# Single-channel and multi-channel electrospinning for the fabrication of PLA/PCL tissue engineering scaffolds: comparative study of the materials physicochemical and biological properties

Semen Goreninskii [1,2*], Ulyana Chernova [1], Elisaveta Prosetskaya [1], Alina Laushkina [1], Alexander Mishanin [3], Alexey Golovkin [3], Evgeny Bolbasov [1,4*]

[1] *B.P. Weinberg Research Center, Tomsk Polytechnic University, 634000, Tomsk, Lenina av., 30, Russian Federation*

[2] *Onconanotheranostics Laboratory, Shemyakin-Ovchinnikov Institute of Bioorganic Chemistry RAS, 117997, Moscow, Miklukho-Maklaya, st., 16/10, Russian Federation*

[3] *Almazov National Medical Research Center, 197341, St. Petersburg, Akkuratova st., 2, Russian Federation*

[4] *Microwave Photonics Lab, V.E. Zuev Institute of Atmospheric Optics Siberian Branch of the Russian Academy of Sciences, 634055, Tomsk, Academician Zuev sq., 1, Russian Federation*

* **E-mail**: sig1@tpu.ru (S.G.), ftoroplast@tpu.ru (E.B.)

**Abstract**

Fabrication of tissue engineering scaffolds with tailored physicochemical and biological characteristics is a relevant task in biomedical engineering. The present work was focused at the evaluation of the effect of fabrication approach (single-channel or multi-channel electrospinning) on the properties of the fabricated poly(lactic acid)(PLA)/poly(ε-caprolactone)(PCL) scaffolds with various polymer mass ratios (1/0, 2/1, 1/1, 1/2, and 0/1). The scaffolds with same morphology (regardless of electrospinning variant) were fabricated and characterized using SEM, water contact angle measurement, FTIR, XRD, tensile testing and in vitro experiment with multipotent mesenchymal stem cells. It was demonstrated, that multi-channel electrospinning prevents intermolecular interactions between the polymer components of the scaffold, preserving their crystal structure, what affects the mechanical characteristics of the scaffold (particularly, leads to 2-fold difference in elongation). Better adhesion of multipotent mesenchymal stem cells on the surface of the scaffolds fabricated using multichannel electrospinning was demonstrated.

# 1. Introduction

Electrospinning is a unique technology for the fabrication of fibers with nano- and micrometer diameters [1]. Electrospun materials possess high porosity and specific surface as well as controllable mechanical and morphological properties. Owing to these features, electrospun materials are actively searched in wide variety of applications: packaging [2], filtration [3], catalysis [4], soft electronics [5] and tissue engineering [6].

In the field of tissue engineering, electrospun materials are applied as scaffolds, fibrous polymer structures providing the conditions for cell growth and proliferation. Due to the variety of tissues and organs, the task of tailoring of physicochemical and biological characteristics of tissue engineering scaffolds is relevant [7]. Fabrication of composite materials, which combine the properties of two and more components is a promising approach for that purpose [8].

Biodegradable polyesters are widely used for the development of tissue engineering scaffolds [9]. Among them, poly(lactic acid) and poly(ε-caprolactone) may be highlighted: both polymers possess high biocompatibility, approved by FDA for the development of biomedical devices and relatively available. With that, some characteristics of these polymers are polar: PCL demonstrates lower degradation rate, higher hydrophobicity, crystallinity and elongation compared to PLA [10]. For that reason, the development of PLA/PCL-based tissue engineering scaffolds became relevant. Herrero-Herrero et al. fabricated the scaffolds made of PLA, PCL and their mixture in 1/1 mass ratio and studied the effect of the average fiber diameter on the interactions of mesenchymal stem cells with the fabricated scaffolds. The composite scaffolds enhanced chondrogenic differentiation of the cells and were suggested for regeneration of cartilage tissue [11]. Zhang et al. used multi-spinneret technique for the fabrication of scaffolds made of PLA, PCL and their mixtures in mass ratios of 2/1 and 1/2. The results of in vitro experiment demonstrated the enhanced growth of epithelial cells with the increase of PCL content in the materials [12]. Recently, the scaffolds made of PLA, PLLA, PCL and their mixtures (in ratios of 20/80 and 50/50) were fabricated by Oztemur et al. from common spinning solutions in mixture of chlorophorm, ethanol and acetic acid or chlorophorm and acetone (in case of PLA scaffold). The best adhesion of fibroblasts and endothelial cells was observed on composite PLA/PCL scaffold with polymer ratio of 20/80 [13]. Meng et al. fabricated PLA/PCL scaffolds with developed surface of the fibers. Compared to the scaffolds with smooth fiber surface, better viability and adhesion of human osteoblast-like cells was observed. The authors suggested PLA/PCL scaffolds for bone tissue engineering [14]. Scaffaro et al. fabricated PLA/PCL electrospun scaffolds with various polymer ratios (1/0, 3/1, 1/1, 1/3 and 0/1) by supplying the polymers solutions via two separate channels. The authors demonstrated the possibility to vary scaffold strength and hydrolytic degradation rate by changing polymer ratio [15]. Finally, the scaffolds with equal fiber diameter made of PLA, PCL, their copolymer and mixture were fabricated by our group. The best adhesion of multipotent mesenchymal stem cells was observed on the PLA and PCL scaffolds [16].

Thus, two electrospinning variants may be used for the fabrication of PLA/PCL scaffolds: single-channel (the scaffolds are fabricated from common solution) and multi-channel (the polymer solutions are supplied via separate channels). With that, the direct comparison of the physicochemical and biological properties of PLA/PCL scaffolds fabricated using single-channel and multi-channel electrospinning was not performed, yet. As the problem of production of tissue engineering scaffolds with tailored properties plays the key role in biomedical engineering, the aim of the present research was to investigate the effect of electrospinning variant on physicochemical and biological properties of PLA/PCL tissue engineering scaffolds.

## 2. Experimental section

### 2.1 Materials

Poly(lactic acid) (Revode 101, $M_n = 10^4$ Da, Zhejiang Hisun Biomaterials) and poly(ε-caprolactone) ($M_n = 8 \times 10^4$ Da, Sigma) were used for the scaffolds fabrication. The polymers were dissolved in 1,1,1,3,3,3-hexafluoropropan-2-ol (HFIP) (99 %, P&M Invest).

### 2.2 Preparation of the spinning solutions

The scaffolds were fabricated from 7 wt. % spinning solutions of the polymers and their mixtures in HFIP. The solutions were prepared in sealed glass reactor under room temperature and agitation during 24 h until the homogenous solution was obtained. For the preparation of the solutions of PLA/PCL mixtures, the prepared polymer solutions were mixed in needed mass ratios (2/1, 1/1 and 1/2) under room temperature and agitation during 24 h until the homogenous solution was obtained.

### 2.3 Dynamic viscosity measurements

The dynamic viscosity of the prepared spinning solutions was measured using SV-10 viscosimeter (A&D, Japan). Three measurements were performed for each solution.

### 2.4 Scaffolds fabrication

The scaffolds were fabricated using electrospinning setup with three independent channels of polymer solution supply and common collector (⌀=10 mm) developed and assembled in the Tomsk Polytechnic University (TPU, Russia). PLA scaffolds were fabricated under applied voltage of 37.5 kV, collector-to-needle distance of 80 mm and solution flow rate of 3 mL/h. The parameters for PCL scaffold fabrication were set as follows: applied voltage of 30 kV, collector-to-needle distance of 140 mm and flow rate of 3 mL/h. For the fabrication of PLA/PCL scaffolds using multi-channel electrospinning, the following configurations were used: 2 channels of PLA solution supply and 1 channel of PCL solution supply; 1 channel of PLA solution supply and 1 channel of PCL solution supply; and 1 channel of PLA solution supply and 2 channels of PCL solution supply (designated as 2/1M, 1/1M and 1/2 M, respectively). 6 mL of the spinning solutions were used for the fabrication of each scaffold.

## 2.5 Scanning electron microscopy

Scanning electron microscopy of the fabricated scaffolds was performed using Quanta 200 3D (FEI, USA) microscope in high vacuum mode at an accelerated voltage of 20 kV. Before the microscopy, the samples were coated with thin layer of gold using SmartCoater sputtering system (Jeol, Japan). The average diameter of the scaffolds fibers was calculated from not less than 50 measurements using ImageJ software (National Institute of Health, USA).

## 2.6 Porosity measurement

The porosity of the fabricated scaffolds was investigated using liquid intrusion method adapted from [17]. To do that, the scaffold samples with size of 10×10 mm were weighted using ALC-210d4 analytical balance (Acculab, USA) and placed in 2 mL of 96% ethanol (Sigma) for 24 h. Then, the samples were taken form ethanol and weighted again. The scaffolds porosity was calculated using the formula (1):

$$\varepsilon = \frac{(m_2-m_1)/\rho_{Ethanol}}{(m_2-m_1)/\rho_{Ethanol}+m_1/\rho_{polymer}} \times 100\% \quad (1), \text{where}$$

$m_1$ is sample mass before exposure to ethanol, $m_2$ is sample mass after exposure, $\rho_{ethanol}$ is the density of ethanol (0.807 g/mL), and $\rho_{polymer}$ is the density of PLA and PCL (1.250 and 1.145 g/mL, respectively). Three measurements were performed for each sample.

## 2.7 Water contact angle measurement

Water contact angle of the fabricated scaffolds was measured using EasyDrop DSA20 installation (Kruss, Germany). To do that, a drop of distilled water with 3 µL volume was placed on the sample surface. Five measurements were performed for each sample.

## 2.8 Fourier-transform infrared spectroscopy (FTIR)

FTIR spectra of the fabricated scaffolds in the range of 1000-2000 cm$^{-1}$ were recorded using Cary 630 spectrometer (Agilent, USA).

## 2.9 X-ray diffraction (XRD)

XRD patterns of the fabricated scaffolds in the range of 5-40° were recorded using XRD-6000 diffractometer (Shimadzu, Japan). The size of PCL crystallites in the fabricated scaffolds was calculated using Scherrer equation (2):

$$l_c = \frac{\kappa \lambda}{cos\theta \beta} \quad (2), \text{ where}$$

λ is the wavelength of the incident radiation (Cu K-alpha, λ = 1.54056 Å), β is the width of the reflection at a half height (FWHM), θ is the angle of the diffraction, and k = 0.9. The estimation of the FWHM parameter was performed by interpolating the X-ray pattern using a Gauss function in Origin 2021 software (Origin Lab, USA).

## 2.10 Tensile testing

Tensile testing of the fabricated scaffolds was performed using Instron 3344 testing machine (Instron, USA) in accordance with ISO 9073-3:1989 standard "Textiles — Test methods for nonwovens — Part 3: Determination of tensile strength and elongation". Five measurements were performed for each sample.

## 2.11 Cell studies

Biocompatibility of the fabricated scaffolds was evaluated using multipotent mesenchymal stem cells (MMSC) according to the protocol published earlier [16,18,19]. The cells were cultured in CO$_2$-incubator at a temperature of 37°C and 5% CO$_2$ content in alpha-MEM medium doped with 10% of fetal bovine serum, 1% L-glutamine and 1% solution of penicillin/streptomycin.

The rectangular samples with size of 12×8 mm were soaked for 30 min in phosphate buffered saline (PBS) doped with 2% of penicillin/streptomycin solution followed by triplicate washing with PBS. The samples were placed into the wells 24-well culture plate, 1 mL of the culture medium was added to each sample and the samples were incubated in $CO_2$-incubator for 24 hours. Then, the medium was removed and cells suspension (50000 cells/mL) was added to each sample. The cells were cultured with the samples for 72 h in $CO_2$-incubator. The experiment was performed in triplicates.

After 72 h of cultivation, the samples were transferred to the wells of the new culture plate, washed with PBS and treated with 4% paraformaldehyde for 10 min. Then, the samples were treated with 0.05 % solution of Triton X-100 during 3 min followed by triplicate washing with PBS. The solution of rhodamine-conjugated phalloidin (ThermoFisher Scientific, Massachusetts, USA) in 1 % fetal bovine serum in PBS (1:500 dilution) was added to each well and incubated for 20 min at a room temperature. Then, the samples were washed with PBS fivefold. After that, the solution of 4,6-diamidino-2-phenylindole (DAPI, dilution 1:40,000, Sigma-Aldrich, USA) was added in each well for 40 s and the samples were intensively washed in PBS.

Inverted fluorescence microscope Axiovert (Zeiss, Germany) equipped with Canon camera was used for cells visualization. The number of adhered cells was determined from the images with ×10 magnification, while the sample area covered with cells was calculated from the images with ×40 magnification.

## 2.12 Statistics

Statistical analysis of the results was performed using Prism 9 software (GraphPad, USA).

# 3. Results and discussion

## 3.1 Optimization of PLA/PCL composite scaffolds fabrication using single-channel electrospinning

Viscosity of the spinning solution considerably determines the parameters of fabrication and morphological features of the resulting scaffold. For that reason, in order to optimize the parameters of PLA/PCL scaffolds fabrication using single-channel electrospinning, the viscosity of the spinning solutions was measured (Table 1).

Table 1. Dynamic viscosity of the solutions of PLA, PCL and their mixtures.

| PLA/PCL mass ratio in the spinning solution | Dynamic viscosity, Pa×s |
| --- | --- |
| 1/0 | 0.39±0.02 |
| 2/1 | 0.75±0.17 |
| 1/1 | 0.98±0.13 |
| 1/2 | 1.90±0.28 |
| 0/1 | 3.15±0.22 |

Dynamic viscosity of the PLA solution was found at 0.39±0.02 Pa×s, while for PCL solution it was by about an order of magnitude higher (3.15±0.22 Pa×s). The viscosity of the solutions of PLA/PCL mixtures increased with the increase of PCL content (Table 1). According to the results of the spinning solutions viscosity study, connector-to-needle distance in the range of 80-140 mm and applied voltage in the range of 30-37.5 kV were tested for the optimization of PLA/PCL scaffolds fabrication using single-channel electrospinning. Based on the absence of statistically significant differences in average fiber diameter and its distribution, the following single-channel electrospinning parameters were selected for the fabrication of PLA/PCL scaffolds: applied voltage of 37.5 kV and collector-to-needle distance of 80 mm for PLA/PCL scaffold with 2/1 mass polymer ratio (designated as 2/1S), applied voltage of 30 kV and collector-to-needle distance of 140 mm for PLA/PCL scaffold with 1/1 mass polymer ratio (designated as 1/1S) and applied voltage of 37.5 kV and collector-to-needle distance of 80 mm for PLA/PCL scaffold with 1/2 mass polymer ratio (designated as 1/2S).

**3.2 Morphology of the fabricated scaffolds**

Morphological features (particularly, fiber diameter, porosity) affect physicochemical, mechanical and biological properties of tissue engineering scaffolds [20]. For that reason, one of the aims of that study was to select the regimes providing the scaffolds with equal PLA/PCL ratio with the same morphology regardless of the fabrication method (multi-channel or single-channel electrospinning). SEM images and histograms of the fiber diameter distribution of the fabricated scaffolds are presented in Fig. 1.

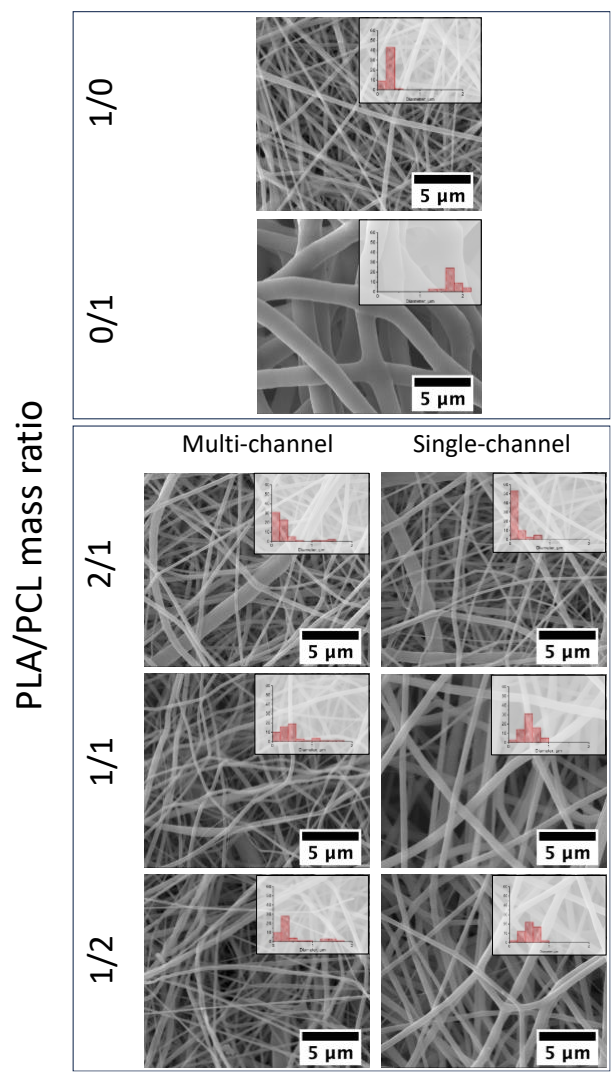

**Fig. 1.** SEM images and histograms of fiber diameter distribution of the PLA/PCL scaffolds with various polymer ratios fabricated using multi-channel and single-channel electrospinning

PLA- and PCL-based scaffolds were formed by cylindrical chaotically interweaved fibers with the diameters of 0.26±0.07 and 1.74±0.19 μm, respectively (Table 2). In case of the scaffolds fabricated using multi-channel electrospinning, the increase of PCL content resulted in the increase of the average fiber diameter due to the thicker PCL fibers, what was also displayed in the fiber diameter histograms by maximum at 1.5 μm and high standard deviation (Table 2, Fig. 1). The selected regimes of single-channel electrospinning allowed to fabricate the scaffolds without statistically significant differences in fiber diameter compared to the scaffolds with the equal

PLA/PCL ratio fabricated using multi-channel electrospinning. Moreover, the porosity of the scaffolds with equal PLA/PCL ratio was the same regardless of fabrication method (Table 2).

Table 2. Average fiber diameter and porosity of PLA/PCL scaffolds with various polymer ratios fabricated using different electrospinning variant

| PLA/PCL ratio | Average fiber diameter, µm | | Porosity, % | |
|---|---|---|---|---|
| | Multi-channel electrospinning | Single-channel electrospinning | Multi-channel electrospinning | Single-channel electrospinning |
| 1/0 | 0.26±0.07 | | 79.2±0.4 | |
| 2/1 | 0.30±0.29 | 0.20±0.18 | 71.5±0.7 | 71.5±0.3 |
| 1/1 | 0.48±0.37 | 0.53±0.19 | 86.4±0.2 | 85.9±1.4 |
| 1/2 | 0.46±0.43 | 0.52±0.17 | 85.0±0.8 | 85.2±1.9 |
| 0/1 | 1.74±0.19 | | 84.0±0.9 | |

\* - p<0.05, statistically significant differences compared to multi-channel electrospinning group (Mann-Whitney test)

### 3.3 Surface hydrophilicity of the fabricated scaffolds

Surface hydrophilicity of the tissue engineering scaffolds significantly determines their interactions with recipient's cells and tissues [21]. The results of the study of the surface hydrophilicity of the fabricated scaffolds are presented in Table 3.

Table 3. Water contact angle of PLA/PCL scaffolds with various polymer ratios fabricated using different electrospinning variants

| PLA/PCL ratio | Water contact angle, ° | |
|---|---|---|
| | Multi-channel electrospinning | Single-channel electrospinning |
| 1/0 | 130.6±3.7 | |
| 2/1 | 129.1±2.3 | 129.6±4.6 |
| 1/1 | 129.5±1.6 | 129.8±1.4 |
| 1/2 | 129.7±2.4 | 129.8±1.6 |
| 0/1 | 131.8±1.2 | |

* - $p<0.05$, statistically significant differences compared to multi-channel electrospinning group (Mann-Whitney test)

All fabricated scaffolds demonstrated hydrophobic character of the surface, which is typical for electrospun PLA/PCL scaffolds [22,23]. However, the absence of statistically significant differences between the composite PLA/PCL scaffolds with equal polymer ratios fabricated using muti- and single-channel electrospinning should be noted. As the chemical composition of the scaffolds (polymer ratio) is equal, morphology significantly affects their wetting performance [24]. Statistically significant differences in water contact angle were not observed regardless of scaffold fabrication technique (Table 3). Thus, the same morphology of the composite PLA/PCL scaffolds with equal polymer ratios was indirectly confirmed.

## 3.4 FTIR spectroscopy of the fabricated scaffolds

FTIR spectroscopy allows investigating the chemical structure of polymer composites and identify intermolecular interactions in them [25]. FTIR spectra of the fabricated scaffolds are presented in Fig. 2.

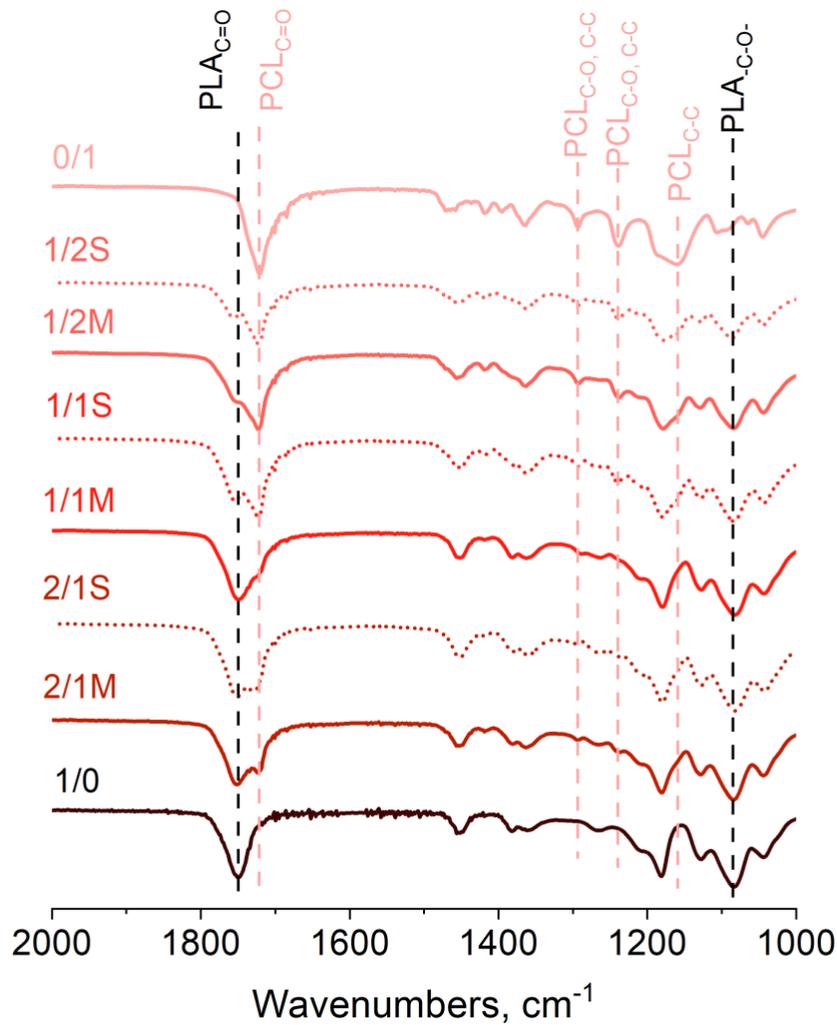

**Fig. 2.** FTIR spectra of PLA/PCL scaffolds with various polymer ratios fabricated using different electrospinning variants

FTIR spectra of the PLA and PCL scaffolds were presented by typical absorption bands of these polymers: at 1181 and 1082 cm$^{-1}$ corresponding to –C–O– stretch and at 1750 cm$^{-1}$ corresponding to –C=O carbonyl stretch for PLA and at 1293 и 1160 cm$^{-1}$ corresponding to C–O and C–C stretching in the crystalline phase, at 1239 cm$^{-1}$ corresponding to –C–O– stretch and at 1721 cm$^{-1}$ corresponding to –C=O carbonyl stretching for PCL (Fig. 2) [26,27]. According to the results of FTIR spectroscopy, fabrication of PLA/PCL scaffolds by multi-channel electrospinning prevented intermolecular interactions between the macromolecules: the shifts of characteristic absorption bands were not observed. These observations are consistent with the results of Scaffaro et al. [15]. On the other hand, on the spectra of PLA/PCL scaffolds fabricated using single-channel electrospinning the shift of the absorption band corresponding to PLA –C=O carbonyl stretching was observed (from 1750 to 1756 cm$^{-1}$). That effect was also observed for PLA/PCL mixtures by Sundar et al. and attributed to the formation of hydrogen bonds between the macromolecules [28]. Thus, it may be concluded that multi-channel electrospinning allows preventing intermolecular interactions between the components of composite scaffold, which may affect its mechanical properties.

**3.4 Crystal structure of the fabricated scaffolds**

Crystal structure of the polymer tissue engineering scaffolds affects their mechanical and even biological characteristics [29]. Crystal structure of the fabricated PLA/PCL scaffolds was investigated using XRD. XRD patterns of the fabricated scaffolds are presented in Fig. 3.

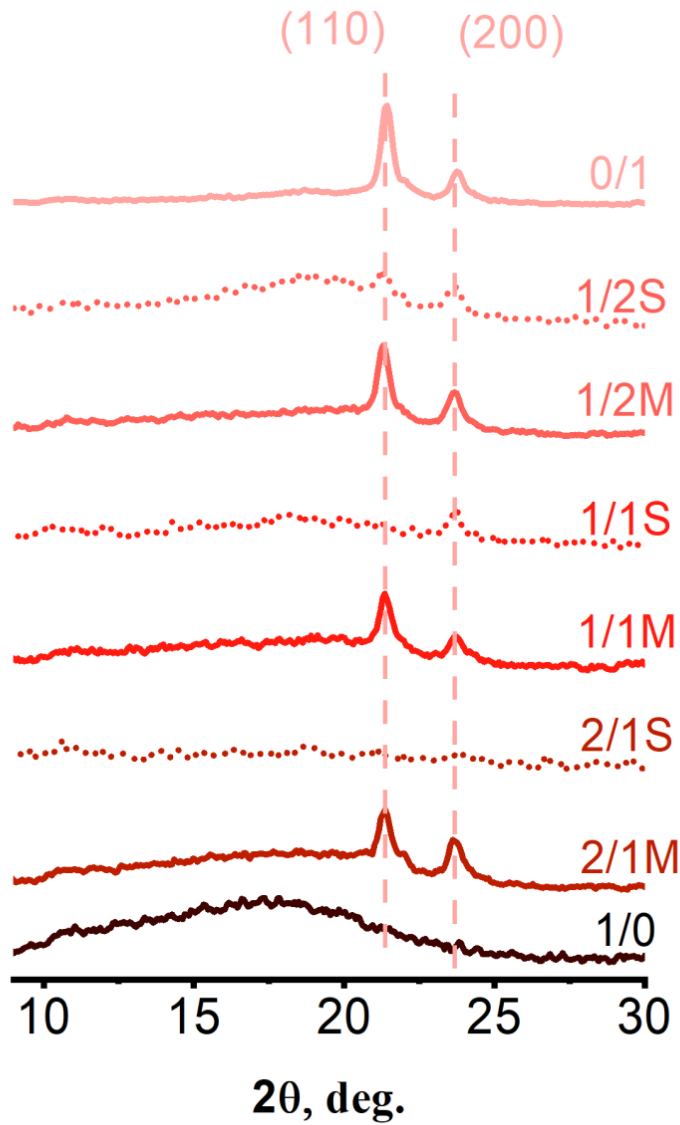

**Fig. 3.** XRD patterns of PLA/PCL scaffolds with various polymer ratios fabricated using multi-channel and single-channel electrospinning

XRD pattern of the PLA scaffold was presented by wide halo, what demonstrates the amorphous structure of the polymer. On the other hand, on XRD pattern of the PCL scaffold two reflexes (at 21.4 and 23.7°) corresponding to (110) and (200) crystal plains of PCL were observed. The recorded diffractograms are typical for PLA and PCL electrospun scaffolds [30]. As during the multi-channel process the components of the composite don't interact with each other (what was demonstrated by FTIR spectroscopy), the XRD pattern of PLA/PCL scaffolds fabricated using multi-channel electrospinning was presented by overlapped patterns of PLA and PCL (Fig. 3). At the same time, XRD patterns of the PLA/PCL scaffolds fabricated using single-channel electrospinning demonstrated the influence of PLA on PCL crystallization. The reflexes of PCL were not observed on XRD pattern of 2/1S sample, what demonstrates its amorphous structure. Single reflex corresponding to (200) crystal plain of PCL was observed on the XRD pattern of 1/1S sample. Finally, both reflexes corresponding to PCL crystal plains were observed. These observations combined with the results of crystallite size calculations (Table 4) demonstrate the hindered crystallization of PCL in PLA/PCL scaffolds fabricated using single-channel electrospinning. Regardless of the scaffold composition, the size of PCL crystallites in the scaffolds fabricated using multi-channel electrospinning was found at 17.5±0.3 nm. On the other hand, the size of PCL crystallites in the 1/1S and 1/2S scaffolds was found around 5 nm. Taking into account the results of FTIR spectroscopy, it may be supposed that PCL crystallization was hindered due to the hydrogen bonding between the macromolecules.

Table 4. PCL crystallite size and mechanical properties of PLA/PCL scaffolds with various polymer ratios fabricated using different electrospinning variants

| PLA/PCL ratio | PCL crystallite size, nm | | Elongation, % | | Young modulus, MPa | |
|---|---|---|---|---|---|---|
| | Multi-channel electrospinning | Single-channel electrospinning | Multi-channel electrospinning | Single-channel electrospinning | Multi-channel electrospinning | Single-channel electrospinning |
| 1/0 | - | | 57±3 | | 330±18 | |
| 2/1 | 17.5±0.3 | - | 83±5 | 176±5* | 354±37 | 308±3 |

| 1/1 | 17.5±0.4 | 4.9±1.5* | 156±51 | 286±15* | 247±5 | 221±35 |
| 1/2 | 17.5±0.3 | 5.3±1.8* | 352±30 | 533±28* | 206±18 | 189±7 |
| 0/1 | 17.5±0.3 | | 521±35 | | 52±6 | |

\* - $p<0.05$, statistically significant differences compared to multi-channel electrospinning group (Mann-Whitney test)

### 3.5 Tensile properties of the fabricated scaffolds

PLA and PCL scaffolds demonstrated opposite mechanical characteristics. PLA scaffolds exhibited higher Young modulus of 330±18 MPa and lower elongation of 57±3 %. Regardless of the fabrication approach, the increase of PCL content in the composite scaffolds resulted in the decrease of Young modulus and increase of elongation (to 189±7 MPa и 533±28 %, respectively). With that, the composite scaffolds fabricated using multi-channel electrospinning demonstrated lower elongation and higher Young modulus (Table 4). As the scaffolds with equal PLA/PCL ratio fabricated using multi-channel and single-channel electrospinning possessed the same morphology, polymers crystallinity is the main factor determining the changes in their mechanical properties. The observed effect is in a good agreement with the study of Flamini et al. [31], where lower elongation and higher elastic modulus were observed for the annealed PCL nanofibers with high crystallinity. Thus, the variation of PLA/PCL mass ratio is an efficient approach for the fabrication of tissue engineering scaffolds with tailored mechanical characteristics. The fabrication method (multi-channel or single-channel electrospinning) determines the mechanical properties of the final scaffold via the changes in components crystallinity and intermolecular interactions.

### 3.6 Cell adhesion

The adhesion of MMSC was significantly affected by the fabrication technique and PLA/PCL ratio in the fabricated scaffolds. The round-shaped and elongated cells were observed on the surface of all samples (Fig. 4).

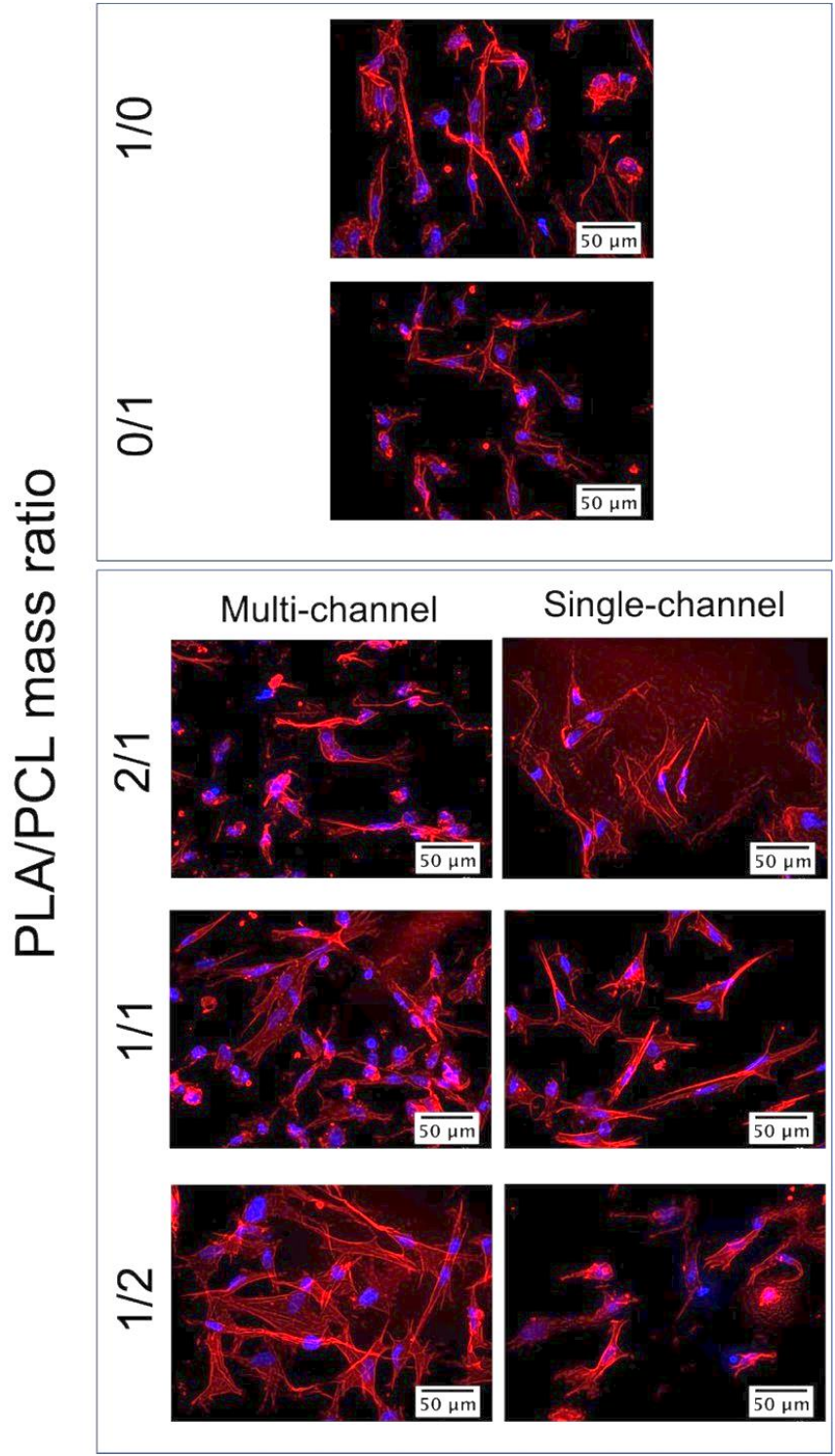

**Fig. 4.** MMSC adhered to the surface of PLA/PCL scaffolds with various polymer ratios fabricated using multi-channel and single-channel electrospinning

Table 5. The number and area of the cells adhered to the surface of PLA/PCL scaffolds with various polymer ratios fabricated using different electrospinning variants

| PLA/PCL ratio | Number of adhered cells, cells/mm$^2$ | | Cell-covered area, % | |
|---|---|---|---|---|
| | Multi-channel electrospinning | Single-channel electrospinning | Multi-channel electrospinning | Single-channel electrospinning |
| 1/0 | 322±53 | | 10.1±2.4 | |
| 2/1 | 228±32 | 46±16* | 9.2±1.2 | 6.7±2.0* |
| 1/1 | 411±80 | 183±29* | 28.2±9.0 | 16.1±3.0* |
| 1/2 | 386±20 | 104±25* | 29.5±5.3 | 6.5±4.2* |
| 0/1 | 236±50 | | 11.4±1.9 | |

\* - p<0.05, statistically significant differences compared to multi-channel electrospinning group (Mann-Whitney test)

Single cells were observed on the surface of PLA and PCL scaffolds in the quantity of 322±53 and 236±50 cells/mm$^2$, respectively, covering 10-12 % of the samples surface. The higher number of cells adhered to the surface of PLA scaffold was observed by our group previously [16]. PLA/PCL ratio in the fabricated scaffolds had non-homogenous effect on the amount of the adhered cells. Regardless of the electrospinning variant, the highest number of adhered cells was observed on the scaffolds with PLA/PCL ratio of 1/1 (Table 5). With that, the number of cells adhered to the surface of the scaffolds fabricated using multi-channel electrospinning, was ≈2-5 times higher compared to the scaffolds fabricated using single-channel variant. Higher cell-covered area was also observed for the scaffolds fabricated using multi-channel electrospinning.

Adhesion of cells to the tissue engineering scaffolds is a complex process, which is affected by the scaffold morphology [11], surface chemistry of the fibers [32], crystal structure of the polymer components [33], mechanical characteristics of the scaffold [34] and other factors. PLA/PCL scaffolds with same chemical composition fabricated in the present study had same morphology. Thus, the enhanced MMSC adhesion to the surface of scaffolds fabricated using multi-channel electrospinning was the result of another factors.

## 4. Conclusions

The physico-chemical and biological properties of PLA/PCL tissue engineering scaffolds fabricated using two electrospinning variants (multi-channel and single-channel) were directly compared. For that, the scaffolds with same polymer ratios and morphology were fabricated. It was shown that multi-channel electrospinning allows preserving the crystal structure of the scaffold components by preventing intermolecular interactions. The preserved crystal structure resulted in lower elongation and higher Young modulus of the PLA/PCL scaffolds fabricated using multi-channel electrospinning. It was also demonstrated that the scaffolds fabricated using multi-channel electrospinning provided significantly better adhesion of multipotent mesenchymal stem cells. Thus, not only chemical composition of the tissue engineering scaffold, but also variant of its fabrication significantly affect its physicochemical and biological properties.

## Acknowledgements

This research was supported by TPU development program Priority 2030 (Priority-2030-NIP/IZ-115-375-2023).

## Conflict of interest

There are no conflicts to declare.

# Reference

[1]  N. Bhardwaj, S.C. Kundu, Electrospinning: A fascinating fiber fabrication technique, Biotechnol Adv., 2010, **28,** 325–347, 10.1016/j.biotechadv.2010.01.004.

[2]  M. Aman Mohammadi, S.M. Hosseini, M. Yousefi, Application of electrospinning technique in development of intelligent food packaging: A short review of recent trends, Food Sci Nutr., 2020, **8**, 4656–4665, 10.1002/fsn3.1781.

[3]  V. V Kadam, L. Wang, R. Padhye, Electrospun nanofibre materials to filter air pollutants – A review, Journal of Industrial Textiles., 2018, **47**, 2253–2280, 10.1177/1528083716676812.

[4]  S. Li, Z. Cui, D. Li, G. Yue, J. Liu, H. Ding, S. Gao, Y. Zhao, N. Wang, Y. Zhao, Hierarchically structured electrospinning nanofibers for catalysis and energy storage, Composites Communications, 2019, **13**, 1–11, 10.1016/j.coco.2019.01.008.

[5]  Y. Wang, T. Yokota, T. Someya, Electrospun nanofiber-based soft electronics, NPG Asia Mater., 2021, **13**, 10.1038/s41427-020-00267-8.

[6]  M. Rahmati, D.K. Mills, A.M. Urbanska, M.R. Saeb, J.R. Venugopal, S. Ramakrishna, M. Mozafari, Electrospinning for tissue engineering applications, Prog Mater Sci., 2021, **117**, 100721, 10.1016/j.pmatsci.2020.100721.

[7]  E. Tanzli, A. Ehrmann, Electrospun Nanofibrous Membranes for Tissue Engineering and Cell Growth, Applied Sciences, 2021, **11**, 6929, 10.3390/app11156929.

[8]  Z. Mohammadalizadeh, E. Bahremandi-Toloue, S. Karbasi, Synthetic-based blended electrospun scaffolds in tissue engineering applications, J Mater Sci., 2022, **57**, 4020–4079, 10.1007/s10853-021-06826-w.

[9]  Y. Zhang, X. Liu, L. Zeng, J. Zhang, J. Zuo, J. Zou, J. Ding, X. Chen, Polymer Fiber Scaffolds for Bone and Cartilage Tissue Engineering, Adv Funct Mater., 2019, **29**,

10.1002/adfm.201903279.

[10] I. Manavitehrani, A. Fathi, H. Badr, S. Daly, A. Negahi Shirazi, F. Dehghani, Biomedical Applications of Biodegradable Polyesters, Polymers (Basel), 2016, **8**, 10.3390/polym8010020.

[11] M. Herrero-Herrero, S. Alberdi-Torres, M.L. González-Fernández, G. Vilariño-Feltrer, J.C. Rodríguez-Hernández, A. Vallés-Lluch, V. Villar-Suárez, Influence of chemistry and fiber diameter of electrospun PLA, PCL and their blend membranes, intended as cell supports, on their biological behavior, Polym Test., 2021, **103**, 107364, 10.1016/j.polymertesting.2021.107364.

[12] S. Zhang, D. Yan, L. Zhao, J. Lin, Composite fibrous membrane comprising PLA and PCL fibers for biomedical application, Composites Communications, 2022, **34**, 101268, 10.1016/j.coco.2022.101268.

[13] J. Oztemur, S. Ozdemir, H. Tezcan-Unlu, G. Cecener, H. Sezgin, I. Yalcin-Enis, Investigation of biodegradability and cellular activity of PCL/PLA and PCL/PLLA electrospun webs for tissue engineering applications, Biopolymers, 2023, 10.1002/bip.23564.

[14] C. Meng, D. Tang, X. Liu, J. Meng, W. Wei, R.H. Gong, J. Li, Heterogeneous porous PLLA/PCL fibrous scaffold for bone tissue regeneration, Int J Biol Macromol., 2023, **235**, 123781, 10.1016/j.ijbiomac.2023.123781.

[15] R. Scaffaro, F. Lopresti, L. Botta, Preparation, characterization and hydrolytic degradation of PLA/PCL co-mingled nanofibrous mats prepared via dual-jet electrospinning, Eur Polym J., 2017, **96**, 266–277, 10.1016/j.eurpolymj.2017.09.016.

[16] E. Bolbasov, S. Goreninskii, S. Tverdokhlebov, A. Mishanin, A. Viknianshchuk, D.

Bezuidenhout, A. Golovkin, Comparative Study of the Physical, Topographical and Biological Properties of Electrospinning PCL, PLLA, their Blend and Copolymer Scaffolds, IOP Conf Ser Mater Sci Eng., 2018, **350**, 012012, 10.1088/1757-899X/350/1/012012.

[17] S. Soliman, S. Sant, J.W. Nichol, M. Khabiry, E. Traversa, A. Khademhosseini, Controlling the porosity of fibrous scaffolds by modulating the fiber diameter and packing density, J Biomed Mater Res A., 2011, **96A**, 566–574, 10.1002/jbm.a.33010.

[18] V.A. Petrova, A.K. Khripunov, A.S. Golovkin, A.I. Mishanin, I. V. Gofman, D.P. Romanov, A. V. Migunova, N.A. Arkharova, V. V. Klechkovskaya, Y.A. Skorik, Bacterial Cellulose (Komagataeibacter rhaeticus) Biocomposites and Their Cytocompatibility, Materials, 2020, **13**, 4558, 10.3390/ma13204558.

[19] V.A. Petrova, N. V. Dubashynskaya, I. V. Gofman, A.S. Golovkin, A.I. Mishanin, A.D. Aquino, D. V. Mukhametdinova, A.L. Nikolaeva, E.M. Ivan'kova, A.E. Baranchikov, A. V. Yakimansky, V.K. Ivanov, Y.A. Skorik, Biocomposite films based on chitosan and cerium oxide nanoparticles with promising regenerative potential, Int J Biol Macromol., 2023, **229**, 329–343, 10.1016/j.ijbiomac.2022.12.305.

[20] S. Pisani, R. Dorati, B. Conti, T. Modena, G. Bruni, I. Genta, Design of copolymer PLA-PCL electrospun matrix for biomedical applications, React Funct Polym., 2018, **124**, 77–89, 10.1016/j.reactfunctpolym.2018.01.011.

[21] K. Dave, V.G. Gomes, Interactions at scaffold interfaces: Effect of surface chemistry, structural attributes and bioaffinity, Materials Science and Engineering: C, 2019, **105**, 110078, 10.1016/j.msec.2019.110078.

[22] M.J. Mochane, T.S. Motsoeneng, E.R. Sadiku, T.C. Mokhena, J.S. Sefadi, Morphology and Properties of Electrospun PCL and Its Composites for Medical Applications: A Mini Review,

Applied Sciences, 2019, **9**, 2205, 10.3390/app9112205.

[23] H. Maleki, B. Azimi, S. Ismaeilimoghadam, S. Danti, Poly(lactic acid)-Based Electrospun Fibrous Structures for Biomedical Applications, Applied Sciences, 2022, **12**, 3192, 10.3390/app12063192.

[24] W. Cui, X. Li, S. Zhou, J. Weng, Degradation patterns and surface wettability of electrospun fibrous mats, Polym Degrad Stab., 2008, **93**, 731–738, 10.1016/j.polymdegradstab.2007.12.002.

[25] U. Riaz, S.M. Ashraf, Characterization of Polymer Blends with FTIR Spectroscopy, in: Characterization of Polymer Blends, Wiley, 2014: pp. 625–678.

[26] B. Şengül, N. Dilsiz, Barrier properties of polylactic acid/layered silicate nanocomposites for food contact applications, Polymer Science Series A, 2014, **56**, 896–906, 10.1134/S0965545X14060194.

[27] E.M. Abdelrazek, A.M. Hezma, A. El-khodary, A.M. Elzayat, Spectroscopic studies and thermal properties of PCL/PMMA biopolymer blend, Egyptian Journal of Basic and Applied Sciences, 2016, **3**, 10–15, 10.1016/j.ejbas.2015.06.001.

[28] N. Sundar, P. Keerthana, S.A. Kumar, G.A. Kumar, S. Ghosh, Dual purpose, bio-based polylactic acid (PLA)-polycaprolactone (PCL) blends for coated abrasive and packaging industrial coating applications, Journal of Polymer Research, 2020, **27**, 386, 10.1007/s10965-020-02320-0.

[29] M. Polak, K. Berniak, P.K. Szewczyk, J.E. Karbowniczek, M.M. Marzec, U. Stachewicz, PLLA scaffolds with controlled surface potential and piezoelectricity for enhancing cell adhesion in tissue engineering, Appl Surf Sci., 2023, **621**, 156835, 10.1016/j.apsusc.2023.156835.

[30] Y. Lu, Y.-C. Chen, P.-H. Zhang, Preparation and Characterisation of Polylactic acid (PLA)/Polycaprolactone (PCL) Composite Microfibre Membranes, Fibres and Textiles in Eastern Europe, 2016, **24**, 17–25, 10.5604/12303666.1196607.

[31] M.D. Flamini, T. Lima, K. Corkum, N.J. Alvarez, V. Beachley, Annealing post-drawn polycaprolactone (PCL) nanofibers optimizes crystallinity and molecular alignment and enhances mechanical properties and drug release profiles, Mater Adv., 2022, **3**, 3303–3315, 10.1039/D1MA01183A.

[32] M. Asadian, S. Grande, I. Onyshchenko, R. Morent, H. Declercq, N. De Geyter, A comparative study on pre- and post-production plasma treatments of PCL films and nanofibers for improved cell-material interactions, Appl Surf Sci., 2019, **481**, 1554–1565, 10.1016/j.apsusc.2019.03.224.

[33] N.E. Muzzio, M.A. Pasquale, X. Rios, O. Azzaroni, J. Llop, S.E. Moya, Adsorption and Exchangeability of Fibronectin and Serum Albumin Protein Corona on Annealed Polyelectrolyte Multilayers and Their Consequences on Cell Adhesion, Adv Mater Interfaces., 2019, **6**, 10.1002/admi.201900008.

[34] S. Han, J. Kim, G. Lee, D. Kim, Mechanical Properties of Materials for Stem Cell Differentiation, Adv Biosyst., 2020, **4**, 10.1002/adbi.202000247.